\documentstyle[aps,epsfig,twocolumn]{revtex}
\begin{document}
\draft   
\flushbottom
\twocolumn[
\hsize\textwidth\columnwidth\hsize\csname @twocolumnfalse\endcsname

%%%%%%%%%%%%%%%%%%%%%%%%%%%%
%TCIDATA{TCIstyle=Article/art2.lat,aps,revtex}

\title{Square billiard with a magnetic flux}
\author{R. Narevich$^{1,2}$, R. E. Prange$^1$, and Oleg Zaitsev$^1$}
\address{$^1$Department of Physics, University of Maryland, College Park, Maryland
20742\\
$^2$Department of Physics and Astronomy, University of Kentucky, Lexington,
Kentucky 40506}
\maketitle

\tightenlines
\widetext
\advance\leftskip by 57pt
\advance\rightskip by 57pt
\begin{abstract}
Eigenstates and energy levels of a square quantum billiard in a magnetic
field, or with an Aharonov-Bohm flux line, are found in quasiclassical
approximation, that is, for high enough energy. Explicit formulas for the
energy levels and wavefunctions are found. A number of interesting states
are shown, together with their wavefunctions. Some states are diamagnetic,
others paramagnetic, still others both dia- and paramagnetic. Some states
are strongly localized. Related systems and possible experiments are briefly
mentioned.
\end{abstract}


]

\narrowtext
\tightenlines

\section{Introduction}

The behavior of mobile charged particles confined to some region and also
subjected to a magnetic flux has interested physicists since the discovery
of the Hall effect over a century ago. Quantum effects turned out to be
subtle and surprising, as attested by Landau diamagnetism and the
Aharonov-Bohm effect\cite{abfl}. About twenty years ago, with the advent of
the quantum Hall effect and mesoscopic systems, the two dimensional case
became prominent. At about the same time, the development of the subject of
``quantum chaos'' also focussed interest on such systems as being among the
simplest of the ``Gaussian unitary ensemble'' universality class.

There is thus a long history of work on the confined quantum motion of
charged particles in a magnetic flux. It is remarkable that all previous
workers overlooked the fact that many fundamental cases, some of which have
been extensively studied numerically, can be solved and classified
analytically to good approximation. Moreover, the results have an
interesting and suggestive complexity.

In this paper we obtain good approximate solutions to a couple of simply
posed and well studied problems of this type. A preliminary version has
appeared electronically\cite{loc}. Rather than stress the generality of our
method, we focus on a typical problem: a two dimensional charged particle
confined to a square billiard in a perpendicular magnetic flux. Certain
conditions on the flux are required to justify the approximations, and we
also require the energy of the particle to be large. These approximate
solutions are compared to numerical solutions. The two flux configurations
considered are a uniform flux and an Aharonov-Bohm flux line. It is crucial
that the square billiard is integrable. Systems other than the square to
which our methods apply will be mentioned at the end.

There are many recent research papers in which the basic system studied is a
square or rectangular billiard with a magnetic flux. A number of these are
inspired by the experiments of L\'evy {\em et al.}\cite{Levy,Richter}, which
measure the magnetic susceptibility of a collection of a considerable number
of mesoscopic, two dimensional metallic systems, each approximately a
square. The field is nearly uniform over the square in this case.

In the presence of a magnetic flux, there is the possibility that {\em %
persistent currents} exist. In other words, it is possible that the
equilibrium state has a nontrivial current. Indeed, we find that eigenstates
of the quantum system have interesting current densities. Although the
wavefunction has a fairly simple representation, the currents can be quite
complex. Some states are predominantly paramagnetic, others are
predominantly diamagnetic. Still others may support both strong paramagnetic
and strong diamagnetic currents which in total nearly cancel. However, these
states strongly affected by the field are rather rare and most states have
weak persistent currents.

Ideas from the field of quantum chaos have also motivated much work. Since
the magnetic field intuitively has circular symmetry, which `conflicts' with
the symmetry of the square, one might expect chaos to ensue\cite{chaos}, as
is the case with the Sinai billiard. Another theme of quantum chaos is that
of energy level and wavefunction statistics\cite{berryrob}. These statistics
depend on whether time reversal symmetry (or other antiunitary symmetry) is
in force. A natural way to break time reversal is by a magnetic flux. [The
square with a uniform flux still has an antiunitary symmetry, however.]

Diffraction effects, in which a classical length shorter than the wavelength
becomes important, are much studied in this context. This is obviously the
case for the zero radius Aharonov-Bohm flux line\cite{sieberdif}. The sharp
corners of the square also cause diffractive effects in the presence of a
uniform magnetic field. These effects are much smaller than for the flux
line, of course. We give estimates for the parameter range in which such
diffraction becomes important, although we defer study of these effects.

Our main motivation however, is that we add to the store of solvable
problems, and perhaps suggest some experiments. In the textbooks, there are
relatively few such integrable problems, basically, only those which reduce
to one dimension, or separate into several one-dimensional problems. The
square without a magnetic flux is such a case in which the $x$ and $y$
motion separates.

The traditional way to widen the class of approximately solvable problems is
perturbation theory and indeed, our approach is a form of quantum
perturbation theory. We are able to study systems which are integrable
except for a `classically weak' perturbation. Of course, treating weak
perturbations classically is challenging, because the long time behavior may
be chaotic. However, quantum perturbation theory is better behaved and
depends on the short time rather than the long time classical behavior.

On the other hand, perturbations small in this sense can give rise to very
large quantum effects, especially on the wave functions. Moreover, even
perturbations which change some length scale by an amount $\delta L<<\lambda
,$ where $\lambda $ is the wavelength, can have big effects. In terms of the
standard nomenclature, a {\em degenerate} perturbation theory is required,
and a number of unperturbed states are strongly mixed together to give the
final result. Usually, this is done by diagonalization of a small matrix,
but in our case the `matrix' can be quite large. However, rather than just
diagonalizing some matrix by a computer calculation, we obtain the result by
an intuitively appealing Schr\"odinger differential equation.

Having solved, for the first time, this interesting class of problems, we
discovered some other methods of obtaining the solution at the same level of
approximation. We shall present these methods elsewhere\cite{BO}.

\section{Square in a uniform field}

We begin with the case of uniform field. For given velocity $v$, the
cyclotron radius is $R_c=v/\omega _c=cp/eB$ where $\omega _c=eB/mc,$ and $%
p=mv.$ The momentum $p=\hbar k=h/\lambda $ is quantally related to
wavenumber $k$ and wavelength $\lambda $. We define the {\em classical}
small parameter $\epsilon =L/R_c=eBL/\,\hbar ck=2\pi \phi /\phi _0kL$. Here $%
L$ is the length of the side of the square, $\phi $ is the magnetic flux $%
BL^2$ and $\phi _0\,$is the flux quantum $hc/e.$ Small $\epsilon $ allows us
to approximate orbits within the square as straight lines, to first
approximation. This is sometimes known as the Aharonov-Bohm regime\cite{Levy}%
, since the leading quantum effects come from the phase interference effects
associated with the vector potential, and do not depend on the change of
classical orbital motion caused by the Lorentz force. Many potential
experiments are in this parameter range.

We choose units such that the dimensionless field is $B\equiv 2\pi \phi
/\phi _0,$ i.e., $2\pi $ times the number of flux quanta in the square. We
take $L,\,\hbar $ and $2m\ $to be unity so that 
\begin{equation}
\epsilon \equiv B/k<<1.  \label{eps}
\end{equation}
The dimensionless wavenumber $k$ is the number of wavelengths in a side of
the square, up to a factor of $2\pi .$ It satisfies 
\begin{equation}
k>>1,  \label{k}
\end{equation}
which is the basis for the quasiclassical approximation.

We shall see that the condition for standard quantum perturbation theory to
work is 
\begin{equation}
k\sqrt{\epsilon }<<1,  \label{QPB}
\end{equation}
or, in other words, $\sqrt{kB}<<1.$ This is not completely obvious, and in
other contexts\cite{Borgo}, it has been guessed incorrectly that the quantum
`perturbation border' is $k\epsilon <<1,$ i.e. $B<<1,$ which has the simple
meaning that the number of flux quanta in the square is small. We, however,
find that nothing much changes at the border $B\sim 1.$

There may also be a `high energy' condition in the form of a requirement
that $k\epsilon ^b<<1.$ The exponent $b$ depends on the smoothness of the
perturbation. We find that for the uniform field, $b=2$, while for the ideal
flux line $b=1.$ Note that for fixed $B$ the energy can be arbitrarily high,
but if instead $\epsilon $ is kept fixed, there is a limitation on the
energy.

This high energy condition is basically the requirement that diffraction
effects not be too important. From a semiclassical perspective, diffraction
effects occur where the classical system has a length scale as short or
shorter than $\lambda .$ An ideal flux line obviously gives rises to
diffraction effects. Billiards also have such a length scale of course,
namely the distance it takes the confining potential to change from zero
inside the billiard, to infinity outside the billiard. This can be taken
into account by a `Maslov phase' of $\pi $ at the boundary, however. There
are also the sharp corners of the square. The square corners whose angular
opening is $\pi /N,$ where $N=2$ is an integer, are a special case at which
no diffraction occurs\cite{cornerdiff}. With such an angle, the billiard can
be extended by reflection, and the corner in effect disappears. However, in
the presence of a magnetic field, this reflection technique does not work,
and with sufficiently large field, orbits which hit directly into the corner
eventually become important to the semiclassics.

\section{Bogomolny's Quasiclassical Surface of Section Method}

Our approach\cite{pnz} utilizes the quasiclassical surface of section [SS]
method of Bogomolny\cite{bogolss}. Poincar\'e's surface of section is a
surface in classical phase space through which all interesting orbits
repeatedly pass. For two dimensional systems, the surface of section is a
two dimensional phase space. For a billiard, the Birkhoff surface of section
is often chosen. Namely, the space part of the surface of section represents
a point on the boundary at which the orbit bounces and is usually measured
by the distance along the boundary of the billiard. The variable conjugate
to this is the component of momentum parallel to the boundary at the moment
of contact. However, many possible surfaces of section can be considered,
and some are more convenient than others.

Bogolmony's method is a generalization of the ``boundary integral method''%
\cite{bogolss,HornS}, applicable for billiards, and based on Birkhoff's
surface of section, to much more general systems and surfaces of section.
The boundary integral method introduces an operator $K(x,x^{\prime },E)$ and
an integral equation $\psi (x)=\int dx^{\prime }K(x,x^{\prime };E)\psi
(x^{\prime })$. This exact equation has nontrivial solutions only when $E$
is on the spectrum. The SS wavefunction $\psi (x)$ is the normal derivative
of the full wavefunction, $\psi (x)=\partial \Psi ({\bf r)/}\partial n,$
when ${\bf r}$ is at the boundary point $x.$ We should mention that only
recently has the boundary integral method been extended to uniform magnetic
fields in the case that $\epsilon $ is of order unity\cite{HornS}.

Bogolmony's operator $T(x,x^{\prime };E)$ is basically the quasiclassical
approximation to $K.$ It thus takes the particle crossing the SS at position 
$x^{\prime }$ to its next crossing at position $x,$ all at energy $E=k^2.\,$
The quasiclassical approximation to the spectrum is determined by the
existence of solutions of $T\psi =\psi .$ If only the spectrum is of
interest, as it has been for many authors, the condition may be expressed as 
$\det \left[ 1-T(E)\right] =0.$

The operator $T$ is given quite generally by 
\begin{equation}
T(x,x^{\prime };E)=\left[ \frac 1{2\pi i}\left| \frac{\partial
^2S(x,x^{\prime };E)}{\partial x\partial x^{\prime }}\right| \right] ^{\frac 
12}\exp \left[ iS(x,x^{\prime };E)\right]  \label{T}
\end{equation}
where $S=\int_{x^{\prime }}^x{\bf p}\cdot d{\bf r}\,${\bf \ }is the action
integral along the classical path from $x^{\prime }$ to $x$. Note that,
rather than giving position and momentum on the SS, positions at two
sequential crossings of the SS are given. It is assumed, for notational
convenience, that there is a unique orbit from $x^{\prime }$ to $x.$ We also
suppress the Maslov phase. Note that $T$ is semiclassically unitary.

The classical action $S(x,x^{\prime })$ generates the {\em surface of
section map.} Namely, the momenta conjugate to $x,x^{\prime }$ are given by 
\begin{equation}
p=\frac{\partial S(x,x^{\prime })}{\partial x};\,\,p^{\prime }=-\frac{%
\partial S(x,x^{\prime })}{\partial x^{\prime }}.  \label{SSM}
\end{equation}
Eq. (\ref{SSM}) implicitly gives the surface of section map $\{p,x\}={\cal M}%
\{p^{\prime },x^{\prime }\}.$

\begin{figure}[tbp]
{\hspace*{0.2cm}\psfig{figure=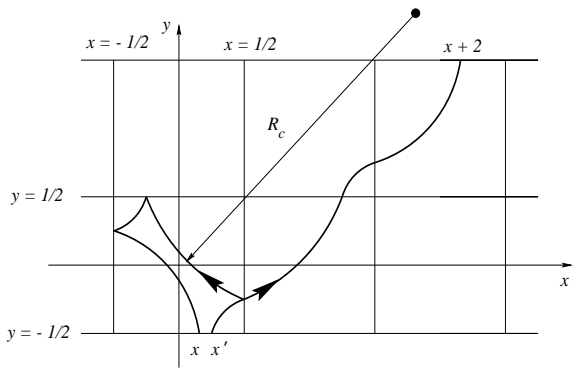,height=7.5cm,width=4.8cm,angle=270}}
{\vspace*{.13in}}
\caption[ty]{ A channel of 2$\times $2 squares replacing the original 1$\times $1
square. Adjacent squares are reflected and the magnetic flux changes sign.
An orbit in the original square is replaced by a nearly straight line orbit
in the channel. The orbit curvature of radius $R_c$ due to the Lorentz force
is exaggerated for clarity in the figure. The orbit shown goes from $%
x^{\prime }$ to $x+2$ in the channel representation and is close to a $(1,1)$
periodic orbit. It is paramagnetic, since the reflections from the square
sides cause it to circulate in the opposite direction from a free orbit in
the field.}
\end{figure}

We attempt to simplify $T$ by astute choice of the surface of section. It
seems simpler to use just one side of the square, rather than all four
sides. It is even easier to use a method of images. Namely, we consider,
instead of a unit square, $x,y\in [-\frac 12,\frac 12]\otimes [-\frac 12,%
\frac 12],$ an infinite channel of width 2 obtained by reflecting the
original square first about $x=\frac 12$ and then about $y=\frac 12,\;$ and
finally repeating the resulting $2\times 2$ square periodically to $x=\pm
\infty $. The flux changes sign in neighboring squares. This geometry is
shown in Fig. 1. There are a continuum of channel solutions. The solutions
to the original square are a subset of these, which exist only at certain
quantized energies. This quantization can be carried out in several ways,
one of which is shown below. 

The SS is taken as the axis $y=-\frac 12\,$ which is identified with $y=%
\frac 32.$ Because the field is classically weak, the path used to calculate
the action is approximated by a straight line. We immediately find 
\begin{equation}
S(x,x^{\prime })=k\sqrt{4+\left( x-x^{\prime }\right) ^2}+\Phi (x,x^{\prime
}),  \label{S}
\end{equation}
i. e. the flux free result plus $\Phi =(e/c)\int {\bf A\cdot }d{\bf r}$,
where the integral is done along the straight line path.

Our scheme\cite{pnz} finds solutions of $T\psi =\psi $ by a perturbation
theory. Taking advantage of the fact that $T$ is unitary, we first solve 
\begin{equation}
\int dx^{\prime }T(x,x^{\prime };E)\psi (x^{\prime })=e^{i\omega (k)}\psi
(x),  \label{Tpsi}
\end{equation}
treating $k=\sqrt{E}$ as a parameter, and then find the energies by solving $%
\omega (k)=2\pi n.$ Given $\psi (x),$ a quadrature which can be carried out
quasiclassically yields the full wave function $\Psi (x,y).$ $\,$Details are
given in Appendix A.

\section{Resonances}

Classical and quantum perturbation expansions in powers of $\epsilon $ fail
near resonances, necessitating modifications which introduce $\sqrt{\epsilon 
}$. Classical resonances correspond to periodic orbits of the unperturbed
system. Periodic orbits on the square correspond to straight line orbits in
the channel from ($x^{\prime },-\frac 12$) to ($x=x^{\prime }+2p/q,$ $\frac 3%
2$). Here $q$ is a positive integer and $p$ is a positive or negative
integer relatively prime to $q.$ Negative and positive $p$ are not
equivalent if there is a magnetic flux. Between resonances, or near
resonances with large $p$ and $q,$ ordinary perturbation theory works. See
Ref. \cite{pnz} for a fuller discussion.

We now specialize to the $(\pm 1,1)$ resonances. These are the simplest
resonances depending strongly on the field and, as we shall see, in some
sense dominate the magnetic response.

We look for a solution of Eq. (\ref{Tpsi}) of the form $\psi (x)=e^{i\kappa
x}u_m(x)$ where $\kappa =k\cos 45^{\circ }=k/\sqrt{2}$ and $u_m$ varies much
more slowly than the exponential. The reason for this choice is that the
phase factor $e^{i\kappa x}$ makes the rapidly varying phases in the
integral $\int dx^{\prime }T\psi $ to be stationary at $x^{\prime }=x-2.$
This corresponds to rectangular shaped periodic orbits of the original
square whose sides make angles of $45^{\circ }$ with the $x$ axis. Such
orbits are shown in Fig. 1.

Because $\epsilon $ is small, the phase $\Phi \,$does not greatly influence
the position of the stationary phase and it suffices\cite{pnz} to evaluate $%
\Phi (x,x^{\prime })$ at $\Phi (x,x-2)=\Phi (x+2,x).$ [The accuracy of this
approximation depends on the smoothness of $\Phi ,$ and failure of the
approximation is related to the onset of diffraction effects mentioned
earlier.] $\Phi (x+2,x)$ is obtained by integrating the vector potential
about the closed rectangular loop, and somewhat remarkably is independent of
gauge. The result\cite{pnz}$\,$ is that $u_m$ satisfies the Schr\"odinger
equation 
\begin{equation}
-u_m^{\prime \prime }+V(x)u_m=E_mu_m  \label{um}
\end{equation}
where $V(x)=-k\Phi (x+2,x)/{\cal L}$ and ${\cal L}=\sqrt{8}$ is the length
of the periodic orbits. Thus we convert the phase $\Phi $ to a `potential' $%
V.$ The {\em transverse energy }$E_m$ enters into the total energy of the
eigenstate, according to Eq. (\ref{Enm}) below, and $m$ is one of two
quantum numbers classifying the states.

\section{Transverse potential and periodic orbits}

For a given resonance and surface of section, there is an effective
potential $V$ which determines the functions $u_m$ and the energy $E_m.$ The
resonance classically corresponds to a continuous set of nonisolated
periodic orbits of the integrable problem. Before perturbation, each of
these orbits has the same action. The potential $V$, to leading order, is
proportional to the change of the action under perturbation, calculated
along the unperturbed path. Each such path is labelled by the parameter $x$,
where it crosses the surface of section. Higher order corrections may also
be found\cite{pnz}.

Knowledge of the potential gives much qualitative insight into the problem.
Its minima, [if smooth], are at stable periodic orbits, as a rule, and its
maxima are at the unstable orbits. In that sense, it represents a classical
island chain. Of course, it was known how to quantize states near the stable
periodic orbits, if a harmonic expansion is allowed. However, the states
that can be found with the aid of $V$ are much more general and in
particular the states with energies $E_m$ near or even above the maxima of
the potential can also be found.

In general, isolated unstable periodic orbits do not support wavefunctions,
but rather `scar' them\cite{scar}. In other words, there appears some excess
weight on the wavefunction near the unstable orbit. In the sense of
Feynman's path integral formulation, there are not enough classical paths
`near' the unstable orbit, to build a complete wavefunction. Here `near'
means that the paths are close to the periodic orbit in the sense of being
well approximated by a quadratic expansion about the periodic orbit.

The same is true in the present case, and wavefunctions cannot be built just
from orbits near an unstable periodic orbit. However, because of the small
parameter $\epsilon $, we can approximate well an entire shell of orbits in
the Feynman integral, and express the result in terms of the potential $%
V(x). $ This shell {\em can} support many states which we find. There {\em %
are} states whose energies are near the maxima of $V$ and thus have extra
weight near the unstable periodic orbits.

The interpretation of $u_m$ is that it gives the structure of the
wavefunction `transverse' to the resonant periodic orbits. Along the
periodic orbits, the wavefunction varies rapidly, but transversely, it
varies relatively slowly. The `longitudinal' and transverse motions are
weakly coupled, because $V$ and thus $u_m$ and $E_m$ depend on $k,$ but this
is easy to take into account.

The concept of `transverse' is a little murky in the quantum case, although
there are cases, including the one under study where it can be made more
precise. We shall not dwell on this further in this paper, however.

We also remark that the Schr\"odinger equation (\ref{um}) requires boundary
conditions, in order to pick out the physically interesting solutions. These
boundary conditions come from the properties imposed on the solution by the
physics of the problem, and are usually simplified by symmetries of the
problem.

\section{Uniform field solution}

\subsection{Effective potential}

For the uniform field, the potential is 
\begin{eqnarray}
V(x) &=&-Bk\left( 
%TCIMACRO{\tfrac 12 }
%BeginExpansion
{\textstyle {1 \over 2}}
%EndExpansion
-2x^2\right) /{\cal L};\;\;x\in [-%
%TCIMACRO{\tfrac 12 }
%BeginExpansion
{\textstyle {1 \over 2}}
%EndExpansion
,%
%TCIMACRO{\tfrac 12 }
%BeginExpansion
{\textstyle {1 \over 2}}
%EndExpansion
],  \nonumber \\
V(x) &=&+Bk\left[ 
%TCIMACRO{\tfrac 12 }
%BeginExpansion
{\textstyle {1 \over 2}}
%EndExpansion
-2(x+1)^2\right] /{\cal L};\;\;x\in [-%
%TCIMACRO{\tfrac 32 }
%BeginExpansion
{\textstyle {3 \over 2}}
%EndExpansion
,-%
%TCIMACRO{\tfrac 12 }
%BeginExpansion
{\textstyle {1 \over 2}}
%EndExpansion
],  \nonumber \\
V(x) &=&V(x+2).  \label{Vuf}
\end{eqnarray}
The factor $\frac 12-2x^2$ is simply the area enclosed by a periodic $(1,1)$
resonant orbit in the shape of a rectangle which bounces from the bottom of
the square at $x.$ In Appendix B we give the corresponding potential for
other resonances.

This periodic potential consists of alternating positive and negative
harmonic potential wells of depth $Bk/2{\cal L}.$ At the boundaries $x=\pm 
\frac 12,$ the second derivative of the potential is discontinuous, a fact
which leads to the mentioned diffraction effects at sufficiently large $%
B^2/k.$

For the $(-1,1)$ resonance, whose orbits are time reversed $(1,1)$ orbits, $%
V(x)$ changes sign. This would not be true if $V$ had its origin in a time
reversal invariant perturbation of the square, for example, a small change
of shape. We can include the $(-1,1)$ resonance in the present scheme by
attributing the region $1/2<x<3/2$ to that resonance. This extension of the $%
x$ coordinate is thus similar to use of a `angle' variable, with positive $x$%
-velocity $v_x$ for $x\in [-\frac 12,\frac 12],$ and negative $v_x$ for $%
x\in [\frac 12,\frac 32].$

If $\sqrt{Bk}=k\sqrt{\epsilon \text{ }}$ is small, the potential $V(x)$ can
be treated perturbatively. On the other hand, for sufficiently large $Bk/%
{\cal L}$, Eq. (\ref{um}) will have low lying tight binding harmonic
oscillator type solutions centered at $x=0,$ (if $B>0$),\thinspace with
energies approximately given by 
\begin{equation}
E_m=-%
%TCIMACRO{\tfrac 12 }
%BeginExpansion
{\textstyle {1 \over 2}}
%EndExpansion
Bk/{\cal L}+(m+%
%TCIMACRO{\tfrac 12 }
%BeginExpansion
{\textstyle {1 \over 2}}
%EndExpansion
)\sqrt{8Bk/{\cal L}}.  \label{EmU}
\end{equation}
This formula holds for $m<<\sqrt{Bk/{\cal L}}$. The lowest wavefunction is
approximately $u_0(x)=e^{-\sqrt{Bk/2{\cal L}}x^2}$ which is arbitrarily
narrow at large energy. These states are {\em paramagnetic, }as follows from
the fact that $\partial E_m/\partial B<0.$ This will be seen more clearly
below.

Eq. (\ref{um}) is valid for larger $m.$ Although very simple analytic
answers are not available, the problem is the well known one of a particle
in a one dimensional periodic potential. We shall see below that we need
only consider the boundary conditions $u(x+2)=\pm u(x).$ This simplification
is a consequence of the symmetry of the square, and something slightly more
complicated would be needed for the rectangle.

The solutions to Eq. (\ref{um}) may be put into four classes, $A,B,C,D$.
Class $A$ states are those with `low' energies near the bottom of the well, $%
E_m\sim -\frac 12Bk/{\cal L}$. For these cases $u(x)$ has support only near $%
x=0$, $\pm 2,\pm 4,...$ These localized states are strongly paramagnetic,
that is, the current circulates in the opposite direction from that of the
particle in free space. In this case $dE_m/dB<0.$ [We shall see that the
transverse energy $E_m$ carries nearly all the field dependence of the total
energy of the corresponding two dimensional eigenstates.]

Class $D$ states have energies much greater than the maximum potential
energy, i.e. $E_m>>\frac 12Bk/{\cal L}$. In this case, the magnetic field is
a small perturbation, since the `potential energy' $V(x)$ in Eq. (\ref{um})
is small compared with the `kinetic energy' given by $u^{\prime \prime }.$
These states are weakly diamagnetic. We shall not consider further this
case. Of course, the approximation of expanding about the $(1,1)$ resonance
eventually breaks down, and higher order resonances are eventually involved%
\cite{pnz}.

Class $C$ has total transverse energy near the top of the potential $V(\pm
1),$ that is, $E_m\sim \frac 12Bk/{\cal L}$ and the states are strongly
affected by the magnetic field. Very crudely, they are somewhat localized or
`scarred' near $x=\pm 1,$ since they spend more time in that region. This
means that they are strongly influenced by the $(-1,1)$ resonance. They are
diamagnetic and $dE_m/dB>0.$

States of class $B$ form a transition region between the low-energy
paramagnetic states, and the higher energy diamagnetic ones, i.e. near $%
E_m\sim 0$. These are states strongly affected by the field, but are such
that $dE_m/dB\sim 0.$

\subsection{Quantization}

There are two states with identical energy in the repeated square scheme.
These are $\psi _I=e^{i\kappa x}u_m(x),$ and $\psi _{II}=e^{-i\kappa
x}u_m(x-1).$ [Changing the sign of $\kappa $ is equivalent to changing the
sign of the field, which in turn can be accomplished by replacing $V(x)$ by $%
V(x+1).$]

Rather than finding the eigenvalues by imposing conditions directly on the $%
\psi $'s, as in the Appendix A, we produce the four two-dimensional
solutions $\Psi $ corresponding to $\psi _{I,II}.$ [Each $\psi (x)$ gives
two $\Psi (x,y)$'s because $y$ and $1-y$ in the strip represent the same
point in the original square.] We show elsewhere\cite{BO} that these states
can also be found directly by a Born-Oppenheimer approximation. One of these
states may be written 
\begin{equation}
\Psi _0(x,y)=e^{i\kappa (x+y)}u_m(x-y-%
%TCIMACRO{\tfrac 12 }
%BeginExpansion
{\textstyle {1 \over 2}}
%EndExpansion
)  \label{PsiI}
\end{equation}
The remaining three states, $1,2,3,$ can be obtained by rotations, e.g. $%
\Psi _1(x,y)={\cal R}\Psi _0(x,y)=\Psi _0(y,-x),$ etc. Here ${\cal R}%
:(x,y)\rightarrow (y,-x)$ is the rotation by $90^{\circ }.$ The gauge can be
chosen so that the Hamiltonian is invariant under ${\cal R}$. Therefore, the
symmetry of an eigenstate can be labelled by $r=0,1,2,3$, where the
eigenvalue of ${\cal R}$ is $i^{-r}.$

Thus, an eigenfunction with symmetry $r$ is given by 
\begin{equation}
\Psi _{(r)}(x,y)=\left( \sum_{s=0}^3i^{rs}{\cal R}^s\right) e^{i\kappa
(x+y)}u_m(x-y-%
%TCIMACRO{\tfrac 12 }
%BeginExpansion
{\textstyle {1 \over 2}}
%EndExpansion
).  \label{Psi}
\end{equation}

[The sequence of rapidly varying phase factors is $\{e^{i\kappa
(x+y)},\;e^{i\kappa (-x+y)},\;e^{-i\kappa (x+y)},\;e^{i\kappa (x-y)}\}.$
These in turn are rapidly varying in the $45^{\circ }$ directions of the
sides of the periodic orbits.] In general, a solution of Eq. (\ref{um})
satisfies the boundary condition $u_m(x+2)=e^{i\beta }u_m(x).$ We need to
find the allowed values for $\beta $ and $\kappa $ which will give the
quantized energies. These conditions are obtained by requiring $\Psi
_{(r)}(x,-\frac 12)$ to vanish, corresponding to Dirichlet conditions in the
original problem. If the wavefunction vanishes on the bottom, it will by
symmetry vanish on the boundary of the square.

Clearly, the sum of the two terms [$s=0,3$] in Eq. (\ref{Psi}) which are
proportional to $e^{+i\kappa x}$ must vanish. This implies $%
u_m(-x)=-i^{-3r}e^{-i\kappa }u_m(x).$ The reflection symmetry $V(x)=V(-x)$
allows us to take $u_m(x)=(-1)^mu_m(-x)$. In turn, this allows quantization
of $\kappa $ in the form $\kappa =n\pi /2,$where $n$ is an integer
satisfying certain conditions depending on $r\,$ and $m.$ Similarly the two
terms proportional to $e^{-i\kappa x}$ in Eq. (\ref{Psi}) give the condition 
$e^{i\beta }=(-1)^r.$ The relationship of $n$ to $r$ and $m$ is 
\begin{equation}
n%
%TCIMACRO{\limfunc{mod} }
%BeginExpansion
\mathop{\rm mod}
%EndExpansion
4=\left[ 2(1-m%
%TCIMACRO{\limfunc{mod} }
%BeginExpansion
\mathop{\rm mod}
%EndExpansion
2)+r\right] 
%TCIMACRO{\limfunc{mod} }
%BeginExpansion
\mathop{\rm mod}
%EndExpansion
4.  \label{groupr}
\end{equation}
It is straightforward to find for the eigenwavenumber 
\begin{equation}
k_{n,m}=2\pi n/{\cal L+}E_m/k.  \label{knm}
\end{equation}
Eq. (\ref{knm}) should be solved iteratively. For example, the first
approximation replaces the $k$ dependence of the term $E_m/k$ by $2\pi n/%
{\cal L}.$ Equivalently, the energy 
\begin{equation}
E_{n,m}=4\pi ^2n^2/{\cal L}^2+2E_m.  \label{Enm}
\end{equation}
Note that $E_m$ depends, relatively weakly, on $n$, since the $k$ in formula
Eq.(\ref{EmU}) should be replaced by $2\pi n/{\cal L}$. The dependence of
the total energy on $B\;$comes through the term $E_m.$ Eqs. (\ref{knm}) and (%
\ref{Enm})\thinspace hold for all symmetries and successive values of $n$ at
fixed $m$ cycle through the representations of ${\cal R}$. Note that, since $%
E_m/k<<k$ the wavelength is given approximately by ${\cal L}/n,$ i.e. the
length of the classical orbits is an integer number of wavelengths.

Thus we have an expression for the energies of a class of states, namely the 
$(\pm 1,1)$ resonant states. They are labelled by integer $n$ which
effectively give the number of wavelengths measured along the $(1,1)$
periodic orbits, and by a second integer $m$ which gives the number of
`nodes' `perpendicular' to this orbit. The very low $m$ states could very
well have been found by earlier methods, since they can be obtained by
expansions about the stable periodic orbits. However, these remarkable
states do not seem to have been noticed heretofore.

\subsection{Orbital magnetism}

The $(1,1)$ states just obtained dominate the magnetic orbital
susceptibility in a parameter range appropriate to experiments\cite{Levy}.
The susceptibility for the square is on a scale rather larger than the
Landau diamagnetism. It is of course not necessary to find the states, or
for that matter, their energies, to calculate the susceptibility. That is
because the susceptibility depends only on the density of states smoothed
over an energy width proportional to the temperature. The Gutzwiller or
better, the perturbed Berry-Tabor trace formula\cite{Richter} is designed to
give exactly that quantity in quasiclassical approximation. Nevertheless,
it's interesting and previously unremarked, that a small subset of states
accounts for most of the magnetism.

We start by finding the orbital susceptibility $\chi $ of a system of
noninteracting electrons in a grand canonical ensemble. This is given by $%
\chi =\partial {\cal M}/\partial B$ where the magnetization ${\cal M}%
=-\partial \Omega (T,\mu ,B)/\partial B.$ Here the grand potential is 
\begin{equation}
\Omega (T,\mu ,B)=-k_BT\sum_a\ln \left[ 1+e^{-(E_a-\mu )/k_BT}\right] .
\label{omega}
\end{equation}
The temperature is $T$, $k_B$ is Boltzmann's constant, and $\mu =k_F^2$ is
the chemical potential. The dependence of $\Omega $ on $B$ comes only
because the eigenenergies $E_a$ depend on $B.$ The sum is over {\em all }%
eigenstates labelled by $a$.

We divide the states $a$ into those relatively few whose energies depend
appreciably on the field and the rest. These field dependent states are
exactly the $(1,1)$ states found above, plus possibly states classically
associated with a few other low resonances, e.g. $(1,3)$. The reason for
this is that the $(1,1)$ states enclose the maximum directed area. They also
have the shortest length ${\cal L}$ which we will see plays a role. The even
shorter $(0,1)$ periodic orbits do not enclose any flux in the approximation
of neglecting the curvature of the orbits although at higher fields they
eventually become important.

Thus, we replace $\sum_a=\sum_b+\sum_{n,m}$ and we can neglect the sum $b$
over field independent states. The second sum, over $(1,1)$ resonance
states, has many fewer terms than the first in a given range of energy.
Since $\mu $ is related to the number of particles, it is nearly independent
of $B.$ It is possible to find $\mu =k_F^2+\delta \mu (B)$ and make a
consistent expansion, and that is indeed necessary if an average over a
large number of squares with canonical statistics is done\cite{Richter}.
However, we just want to illustrate how the $(1,1)$ states dominate the
susceptibility, and we will not consider this further average. Then, we may
approximate

\begin{equation}
{\cal M}(T,\mu ,B)=-\sum_{n,m}\frac{\partial E_{n,m}}{\partial B}f_D\left[
E_{n,m}(B)\right] .  \label{Gr}
\end{equation}
and $f_D$ is the Fermi-Dirac distribution function.

Using the Poisson sum formula, replace the sum on $n$ in Eq. (\ref{Gr}) by
an integral over $k$, and do the integral to obtain 
\begin{eqnarray}
{\cal M} &=&-\frac{k_BT}{k_F}\sum_{r,m,s=0}^\infty {\bf \alpha }_m{\cal L}%
\exp \left( -\frac{\omega _rs{\cal L}}{2k_F}\right)  \nonumber \\
&&\times \sin \left[ {\cal L}s\left( k_F-\frac{E_m}{k_F}\right) \right] .
\label{M}
\end{eqnarray}
Here, $\omega _r=\pi (2r+1)k_BT$ is the Matsubara frequency and $\alpha
_m=\partial E_m/\partial B.$ [We have dropped the `leading' term in the
Poisson formula which totally neglects the discrete quantum nature of the
states and which therefore cannot produce a magnetization.] As an example%
\cite{Levy,Richter}, take $k_BT$ ten times the level spacing of all levels,
i.e. $k_BT=20\pi $ in our units. Then, $\omega _0{\cal L}/2\approx 300.$ If $%
k_F\approx 300-600,$ so that the square contains about 2-6$\times $10$^4$
electrons, the exponential suppression will not be too serious for $%
r=0,\,s=1 $. However, larger $r$ or $s$ do not contribute much. [In the
trace formula approach, $s$ gives the number of repetitions of the primitive
periodic orbit and the sum over $r$ is explicitly carried out.]

Eq. (\ref{M}) shows that states with larger ${\cal L},$ i.e. smaller
spacing, are suppressed, exactly as seen from the trace formula in terms of
periodic orbits. It also shows that relatively large field dependence of the
levels, $\alpha _m,$ is important. For the square, the $(1,1)$ states have
the smallest ${\cal L}$ and also the largest $\alpha _m$. The $(2,1)$
resonance does not couple to a small constant field.

It is also seen that if Eq. (\ref{M}) is averaged over many squares of
somewhat different sizes, because of the oscillations of the sine, the
result is much reduced and it is necessary to go to higher order in $\delta
\mu .$

\begin{figure}[tbp]
{\hspace*{0.2cm}\psfig{figure=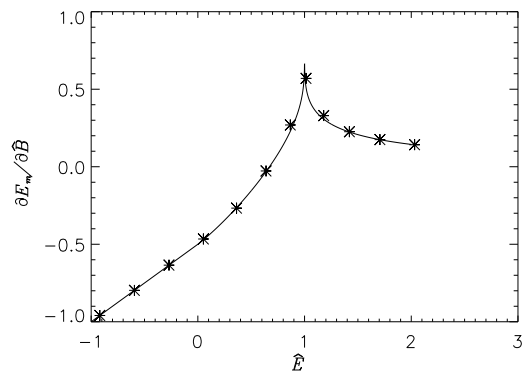,height=5.5cm,width=7.5cm,angle=0}}
{\vspace*{.13in}}
\caption[ty]{ $\partial E_m/\partial \hat B$ vs $E_m/\hat B.$ Stars are for $%
m=0,2,\ldots m_{\max }=14,\ldots $ and $n=62,\,B=25$. The continuous curve  
is Eq. (\ref{qmag}) of the text.}
\end{figure}

In Fig. 2 we show $\partial E_m/\partial \hat B$ as a function of $E_m/\hat B
$ for the (1,1) resonant states, where $\hat B=Bk/2{\cal L}$. According to
Eq. (\ref{M}), if the sign of $\partial E_m/\partial \hat B$ is negative the
contribution of the corresponding state is paramagnetic. For an exact, two
dimensional wavefunction $\Psi _{n,m}$ with energy $E_{n,m}$, it is known
that $\partial E_{n,m}/\partial \hat B$ $=-({\cal L}/k)\int d^2r[{\bf %
r\times j}(x,y)]_z$ , i.e. the expectation value of the $z$-component of the
magnetization density. Here the current is 
\begin{equation}
{\bf j}(x,y)=2%
%TCIMACRO{\limfunc{Re} }
%BeginExpansion
\mathop{\rm Re}
%EndExpansion
\Psi ^{*}(x,y)\left( \frac 1i{\bf \nabla -A}(x,y)\right) \Psi (x,y).
\label{jdef}
\end{equation}

In our approximation, according to Eq. (\ref{Enm}) above, $\partial
E_{n,m}/\partial \hat B$ is equal to $2\partial E_m/\partial \hat B.$ It
follows from Eqs. (\ref{um}) and (\ref{Vuf}) above, that $\partial
E_m/\partial \hat B=\left\langle \hat V(x)\right\rangle _m=\left\langle
u_m\left| \hat V(x)\right| u_m\right\rangle $ where $\hat V=V/\hat B.$ Since 
$\hat V<0$ for $x\in \left[ -\frac 12,\frac 12\right] ,$ and $\hat V>0\,$
for $x\in \left[ \frac 12,\frac 32\right] ,$ etc. we see that the sign of
the magnetic response of a given wavefunction depends on which region of $%
\hat V$ dominates the expectation value. Of course, the classical periodic
orbits in these two regions have the expected sense.

Finally, we may express $\partial E_m/\partial \hat B$ quasiclassically as 
\begin{equation}
\frac{\partial E_m}{\partial \hat B}=\frac{\int dx\hat V(x)\left[ \hat E-%
\hat V(x)\right] ^{-\frac 12}}{\int dx\left[ \hat E-\hat V(x)\right] ^{-%
\frac 12}}.  \label{qmag}
\end{equation}
Here $\hat E=E_m/\hat B$ can be treated as a continuous variable, so that $%
\partial E_m/\partial \hat B$ as a function of $\hat E$ falls on a
continuous curve which in this approximation is independent of $\hat B.$ The
integrals are between the turning points. For $\hat E$ near the minimum $%
\hat V_{\min }$ of $\hat V$ , $\partial E_m/\partial \hat B\approx \hat V%
_{\min }.$ For $\hat E$ at the maximum of $\hat V$, the integrals diverge at 
$x=1$, thus making $\partial E_m/\partial \hat B$ positive, and also giving
the cusp in Fig. 2. Of course, our simple quasiclassical approximation needs
corrections in this case.

The quantization of the transverse motion relates $\hat E=E_m/\hat B\ $and $%
m $ according to the standard formula 
\begin{equation}
\int dx\sqrt{\hat E-\hat V(x)}=\pi (m+\frac 12)/\sqrt{\hat B}.
\label{qquant}
\end{equation}
The spacing of the quantum states therefore depends on $\hat B$. The level $%
m_{top}$ such that $E_{m_{top}}\approx V(1)$ satisfies, approximately, $%
m_{top}\approx 0.6\sqrt{\hat B}.$ The states with $k=142$ and $B=25$ were
chosen in part because with these parameters there is a state very near the
diamagnetic maximum $m_{top}=14,$ and another state with $m=10$ for which $%
\partial E_m/\partial \hat B$ is very small. On the other hand, with these
parameters, while our approximate states are very good for small $m,$ the
transverse momenta are becoming large enough (as compared with the
longitudinal momenta) that our approximation deteriorates significantly at $%
m\geq 10,$ especially in regions of the square where the wavefunction is
small.

\subsection{Visual representations of eigenstates}

Eigenstates of Hamiltonians not satisfying time reversal invariance are
rarely shown graphically in the literature, except for trivial cases. In the
invariant case, numerically produced picture galleries of eigenstates for
systems such as the Bunimovich stadium\cite{MacKauf}, have led to a great
deal of interest and insight, both theoretical\cite{scar} and experimental%
\cite{expscar}.

`Magnetic' states are not real but are inherently complex. A complete
graphical picture of a complex state would seem to require twice the number
of pictures as that necessary for a real state satisfying time reversal
invariance. In addition, the states themselves are gauge dependent, and only
gauge invariant quantities are physically meaningful.

A difference of the magnetic case as compared with the time reversal
invariant case is that generally $\Psi $ or ${\bf j}$ do not vanish along
nodal lines, but rather only at isolated nodal points. There may of course
be symmetries or boundary conditions requiring these quantities to vanish
along a line, but in general the representation of wavefunctions by their
nodal patterns\cite{MacKauf} is not available in the absence of time
reversal invariance.

Two gauge invariant quantities we choose to display are the absolute value
squared of the wavefunction $\left| \Psi (x,y)\right| ^2\,$ and the current
density. The current density is a two dimensional vector field that is
divergence free, ${\bf \nabla \cdot j}=0.$ The one-dimensional surface of
section states are also of interest.

\begin{figure}[tbp]
{\hspace*{1.7cm}\psfig{figure=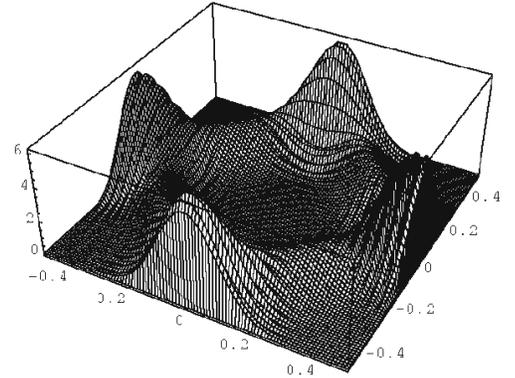,height=6.5cm,width=5.0cm,angle=270}}
{\vspace*{.13in}}
\caption[ty]{ A three dimensional view of the absolute value squared of localized
paramagnetic state for $n=62,\,B=31.4$ and $m=0.$ The cut near the side of
the square is proportional to $u_0$ which is close to a Gaussian. Near the  
side there are interference oscillations from two main terms in Eq. (\ref
{Psi}).}
\end{figure}

We first show a picture of a class $A$ state, which is quite simple to
represent. Fig. 3 shows $\left| \Psi (x,y)\right| ^2,\,$ for $%
n=62,\,k\approx 2\pi 62/{\cal L}\approx 140,\,B=31.4,$ $\sqrt{Bk}\approx 66,$
and $m=0,$ which implies the symmetry $r=0.$ For such a well localized $u_m$%
, each term in Eq. (\ref{Psi}) dominates one side of the rectangular
periodic orbit. For example, near $x=-y=\frac 14$ only the first term, $s=0,$
in the sum (\ref{Psi}) is appreciable. In this region $\left| \Psi
(x,y)\right| \approx $ $u_0(x-y-\frac 12)$ which is well approximated by a
Gaussian.

There is interference near the square edges ({\em e.g. }near ($0,-\frac 12$%
)), and two terms contribute appreciably. Near this point then 
\begin{eqnarray}
&&\Psi (x,y)  \nonumber  \label{side} \\
&\approx &\left| e^{\frac 12n\pi iy}u_0(x-y-%
%TCIMACRO{\tfrac 12 }
%BeginExpansion
{\textstyle {1 \over 2}}
%EndExpansion
)+e^{-\frac 12n\pi iy}u_0(-x-y-%
%TCIMACRO{\tfrac 12 }
%BeginExpansion
{\textstyle {1 \over 2}}
%EndExpansion
)\right|  \nonumber \\
&\approx &\left| 2u_0(x)\cos 
%TCIMACRO{\tfrac 12 }
%BeginExpansion
{\textstyle {1 \over 2}}
%EndExpansion
n\pi y\right|  \label{side}
\end{eqnarray}
so that $\left| \Psi (0,-\frac 12+\frac 1n)\right| $ at its first maximum
near $(0,-\frac 12)$ is about twice as large as $\left| \Psi (\frac 14,-%
\frac 14)\right| $. [Note that $n%
%TCIMACRO{\limfunc{mod}}
%BeginExpansion
\mathop{\rm mod}
%EndExpansion
4=2.$] The current density is thus largest close to the middle of the square
edges and it is of course nearly parallel to the edge there. In this case,
the shape of $\Psi $ close to an edge is given by $u_m(x),$ as can be seen
in Fig. 3.

Fig. 4 shows the streamlines of the current in this state. The direction of
the streamline gives the direction of the current flow, while the density of
streamlines is proportional to the magnitude of the current density. That
is, between any two neighboring streamlines, the same total current flows.
The state in Fig. 4 is {\em paramagnetic,} that is, the current circulates
in the opposite sense from that of a free particle in the field. The choice
of which streamline to display is easily obtained in this case, since each
line crosses a symmetry line like $x=0,\,\,y\in [-\frac 12,0],$ once and
only once. The current has but one interior zero, at the center of the
square.

\begin{figure}[tbp]
{\hspace*{0.2cm}\psfig{figure=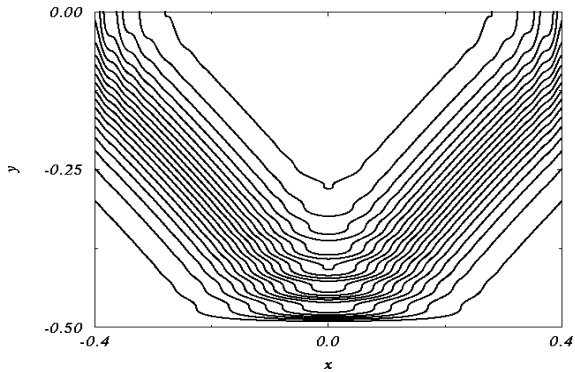,height=7.5cm,width=4.8cm,angle=270}}
{\vspace*{.13in}}
\caption[ty]{ Streamlines of the current of the state shown in Fig. 3. The
interference peaks near $x=0$ are apparent.}
\end{figure}

Fig. 5. shows the state $n=60,$ $B=31.4,$ $m=1$, which has one `transverse
node'. This node does not give rise to a nodal line in the total
wavefunction, of course, although the wavefunction is small in regions
corresponding to the node. The states have rotational symmetry, and we show
a different representation in each corner of the unit square.

One remarkable feature of these states, \thinspace $m=0,1$ is that they are
localized very near the central paramagnetic diamond orbit. Theoretically,
for large $Bk$ this localization can be as tight as one pleases. Although
this is a result of our theory which uses a `potential' function $V(x),$ it
is clear that $V(x)$ is really something that arises from Aharonov-Bohm
phases, and there is no classical localization based on energy
considerations. Although a detailed analysis of this can be a little tricky,
we find the same kind of effect in the AB flux line case, for large $Bk$ and
small $B.$ This localization is thus like Anderson localization\cite
{Anderson} in the sense that, absent phase interference, localization would
not exist. Of course, the random disorder aspects of Anderson localization
are absent.

\begin{figure}[tbp]
{\hspace*{1.5cm}\psfig{figure=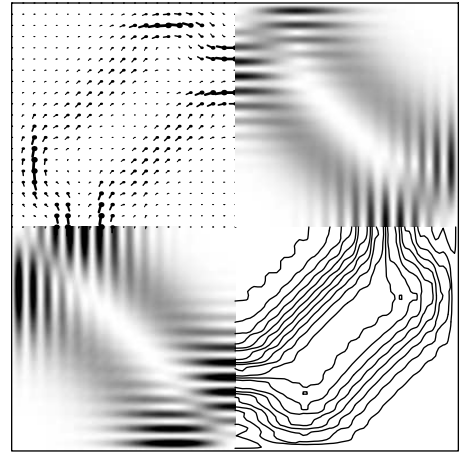,height=6.0cm,width=6.0cm,angle=0}}
{\vspace*{.13in}}
\caption[ty]{ The state with $n=60,B=31.4$ and $m=1.$ Counterclockwise from lower
left, a density plot of $\left| \Psi \right| ^2,$ current streamlines, a
density plot of $\left| {\bf j}\right| ,\,$ and a dot-stick representation 
of the current. [The dot is at the calculated point, the size and direction
of the stick represent the current density magnitude and direction.] The   
function $u_1(x)$ has a node at $x=0.$ This appears as a narrow valley in   
the center of the ridge of current, although there is no node or nodal line
there. The numerical result is shown, which is very well represented by
theory.}
\end{figure}

\begin{figure}[tbp]
{\hspace*{1.5cm}\psfig{figure=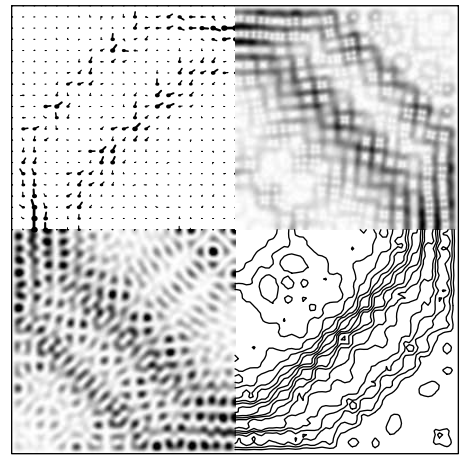,height=6cm,width=6cm,angle=0}}
{\vspace*{.13in}}
\caption[ty]{ The state, $m=14,$ $n=62,B=25$ of maximum diamagnetism, as found
numerically. The same representations as Fig. 5.}
\end{figure}

The next figures, Figs. 6-11, have $n=62,B=25.$ Figs. 6,7 show a state of
class $C$ corresponding to $m=14$ which is energetically at the top of the
periodic potential corresponding to the $(-1,1)$ resonance. This state is
diamagnetic with the current circulating in the opposite sense from the
states with $m=0,1.$ The theoretical and numerical wavefunctions are shown.

The theoretical predictions in this case are considerably less good than for
the states with small $m.$ First, the transverse wavenumber is no longer
quite so small compared with the wavenumber along the path. Second, since
according to Eq. (\ref{Psi}), as many as four approximately found component
states are added, there may be relatively large errors, especially in those
parts of the square where destructive interference is important and the
final wavefunction is small. Nevertheless, Eq. (\ref{Psi}) is a quite good
representation of the more accurate numerical results.

In Fig. 8 we show the state $m=10$, for which $\partial E_{10}/\partial B$
is very small. The streamlines of this state are rather striking. Note that
there are large current loops which have opposite magnetic polarity. Again,
the approximate wavefunction captures many features of the exact one,
although it does not reproduce the finer details. In this case we show also,
in Fig. 9, the transverse state $u_{10}(x)$ and the normal derivative of $%
\Psi _{62,10}$ on the surface of section. Although for small $m,$ the normal
derivative on the surface of section bears an understandable relation to $%
u_m,$ it is quite complicated for transverse energies this large.

Finally, in Fig. 10 we show streamlines for the sequence of states $%
m=6,8,10,12.$ Although there are systematic changes of pattern, we have not
tried to rationalize these changes. We conclude that even though the
wavefunction of Eq. (\ref{Psi}) is fairly simple, it is difficult to foresee
interference patterns when four terms are important.

\begin{figure}[tbp]
{\hspace*{1.5cm}\psfig{figure=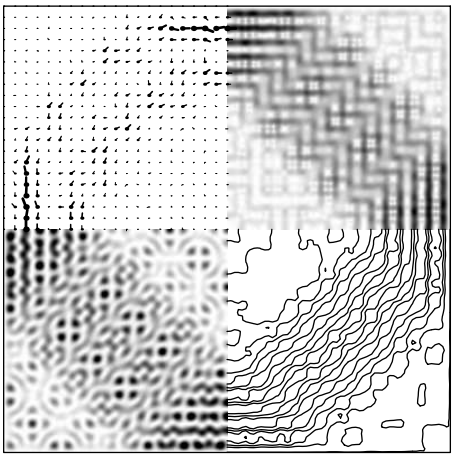,height=6cm,width=6cm,angle=0}}
{\vspace*{.13in}}
\caption[ty]{ The same state as Fig. 6. as predicted by our approximation. For $m$
this large, the theory does relatively poorly, but nevertheless is
qualitatively correct.}
\end{figure}

These relatively complicated states are harder to represent adequately. In
both theory and numerics the ${\bf \nabla \cdot j=}0$ character of the
current is not exact. Numerically following a streamline, particularly in
the neighborhood of a zero of the current is difficult. We therefore imposed
the divergence free character by representing ${\bf j(}x,y){\bf =\nabla
\times \hat z}\chi (x,y).$ We calculated $\chi $ as a symmetrized integral
of $j_x,$ where $j_x$ was obtained either numerically or theoretically. The
streamlines are then contour lines of $\chi $.

The diamagnetic states are not so spatially localized as the low $m$ states,
although they continue to have a sort of localization in `momentum' space,
as we show below. More generally, as $m$ increases, the states become more
delocalized, and eventually become independent of $B.$ This means that for
larger $m$ all four terms in Eq. (\ref{Psi}) make comparable contributions
at an arbitrary typical point $x,y$, whereas in the localized case, only one
or two terms contribute. This gives interference oscillations in $\left|
\psi _m\right| $ near $\left| x\right| \approx \frac 12$ as shown in Fig. 9.

\begin{figure}[tbp]
{\hspace*{1.5cm}\psfig{figure=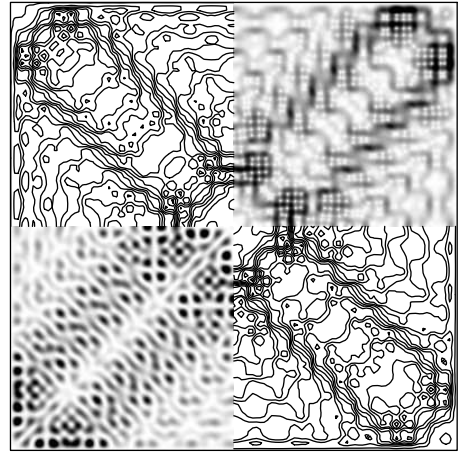,height=6cm,width=6cm,angle=0}}
{\vspace*{.13in}}
\caption[ty]{ The state, $m=10,$ $n=62,B=25$ of nearly cancelling para- and
diamagnetism. Numerical results are shown as before, except for the upper   
left corner where the theoretical streamlines are shown. Theory and
numericals differ primarily in the low current regions. Notice there is a
diamagnetic current loop encircling the diagonals and cancelling
paramagnetic loops in the triangular wedges between the diagonals.}
\end{figure}

\begin{figure}[tbp]
{\hspace*{1.6cm}\psfig{figure=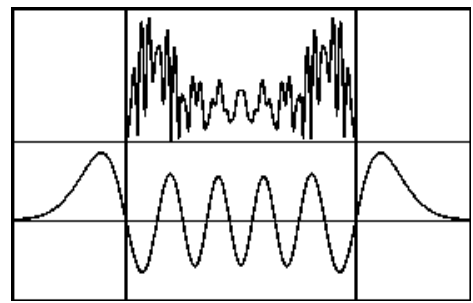,height=3.9cm,width=6.1cm,angle=0}}
{\vspace*{.13in}}
\caption[ty]{ The upper figure shows, for $m=10,$ $n=62,B=25$ the normal
derivative $\left| \partial \Psi _{62,10}/\partial y\right| _{y=-\frac 12},$
obtained numerically. The lower figure is $u_{10}(x).$ Because $u_{10}(x)$
extends significantly outside the domain $\left[ -\frac 12,\frac 12\right] ,$
and is `folded' back into that domain with a rapidly varying additional
phase factor to construct $\partial \Psi /\partial n$, these two constructs
are not so intuitively related as for lower $m$ states which are more
localized.}
\end{figure}

\subsection{Momentum localization}

We have assumed that $n>>m,$ and that $u_m$ is slowly varying compared with $%
e^{i\pi nx/2}.$ We may expand $u_m=\sum \hat u_{m,l}e^{i\pi lx}$, where $l$
is an integer for $r$ even and half odd integer for $r$ odd. Also $\hat u%
_{m,l}=(-1)^m\hat u_{m,-l}.$ The unperturbed states [$\sin \pi p(x+\frac 12%
)\sin \pi q(y+\frac 12)$] can be labelled by integers $p,q$ with unperturbed
energies $(p^2+q^2),$ dropping a factor $\pi ^2.$

Eq. (\ref{Psi}) is a superposition of unperturbed states with quantum
numbers $p=\frac 12n+l,$ $q=\frac 12n-l.$ In particular, the energies $%
p^2+q^2=\frac 12n^2+2l^2$ are closer to the base energy $\frac 12n^2$ than
to the base energy of the next representation, $\epsilon _{n+1}\approx \frac 
12n^2+n.$ Of course, if the perturbation is symmetric under rotation, the
next base energies coupled are $\epsilon _{n\pm 4}\approx \epsilon _n\pm 4n.$

There are, however, other unperturbed states with $p^2+q^2\approx \epsilon
_n.$ For example, $7^2+49^2=\epsilon _{70}.$ However, the matrix elements of
a smoothly perturbed Hamiltonian, ${\cal H}_{pq,p^{\prime }q^{\prime }}$, in
the unperturbed basis, are small if $\left| p-p^{\prime }\right| $ or $%
\left| q-q^{\prime }\right| $ is large. Because the perturbation due to a
uniform field has a singular third derivative, these perturbations drop off
as a power law, and this relatively long range effect in momentum space is
the source of the diffraction corrections.

In Fig. 11 we show the magnitude of the amplitudes of the unperturbed states
combining to make the state $n=62,$ $B=25,$ $m=14.$ The area of a circle is
proportional to the square of the amplitude. Above the main diagonal we show
the theoretical result. All amplitudes lie on the line $31-l,$ $31+l.$
Below, we show the result for the numerical wavefunction. The circle $%
p^2+q^2=2\cdot 31^2$ is also shown. Our theory is starting to need
corrections for $m$ this large.

\begin{figure}[tbp]
{\hspace*{1.5cm}\psfig{figure=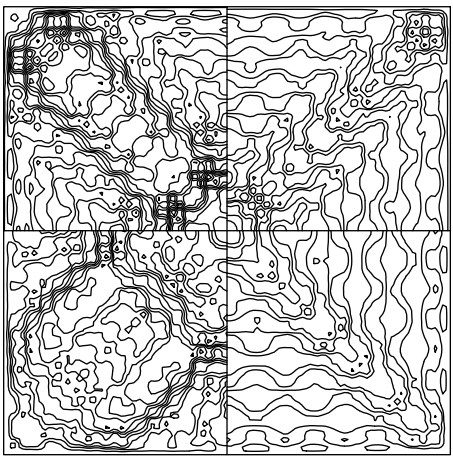,height=6cm,width=6cm,angle=0}}
{\vspace*{.13in}}
\caption[ty]{ Streamlines for a sequence of states, $m=6,8,10,12,$ $n=62,B=25$
counterclockwise from the lower right.}
\end{figure}

Thus, an interpretation of our method which yields the $(1,1)$ resonance
states of Eq. (\ref{Psi}) is that we effectively diagonalize the Hamiltonian
in a basis restricted to the unperturbed states nearly `degenerate' with $%
\epsilon _n$ and close to $\frac 12n,\;\frac 12n.$ This is the case for the
uniform field, and indeed, we achieve agreement between full numerical
diagonalization, diagonalization restricted to `degenerate' states, and the
procedure using the solution of the differential equation Eq. (\ref{um}).

\section{Aharonov-Bohm Flux Line}

The above approach can be generalized to deal with nonuniform flux
configurations. To get the potential $V$ associated with the $(1,1)$
resonance, all that is needed is to be able to calculate the flux contained
within a $(1,1)$ periodic orbit. We consider here the case of the
Aharonov-Bohm flux line [ABFL]. Some further results are published elsewhere%
\cite{dresAB}.

\begin{figure}[tbp]
{\hspace*{1.0cm}\psfig{figure=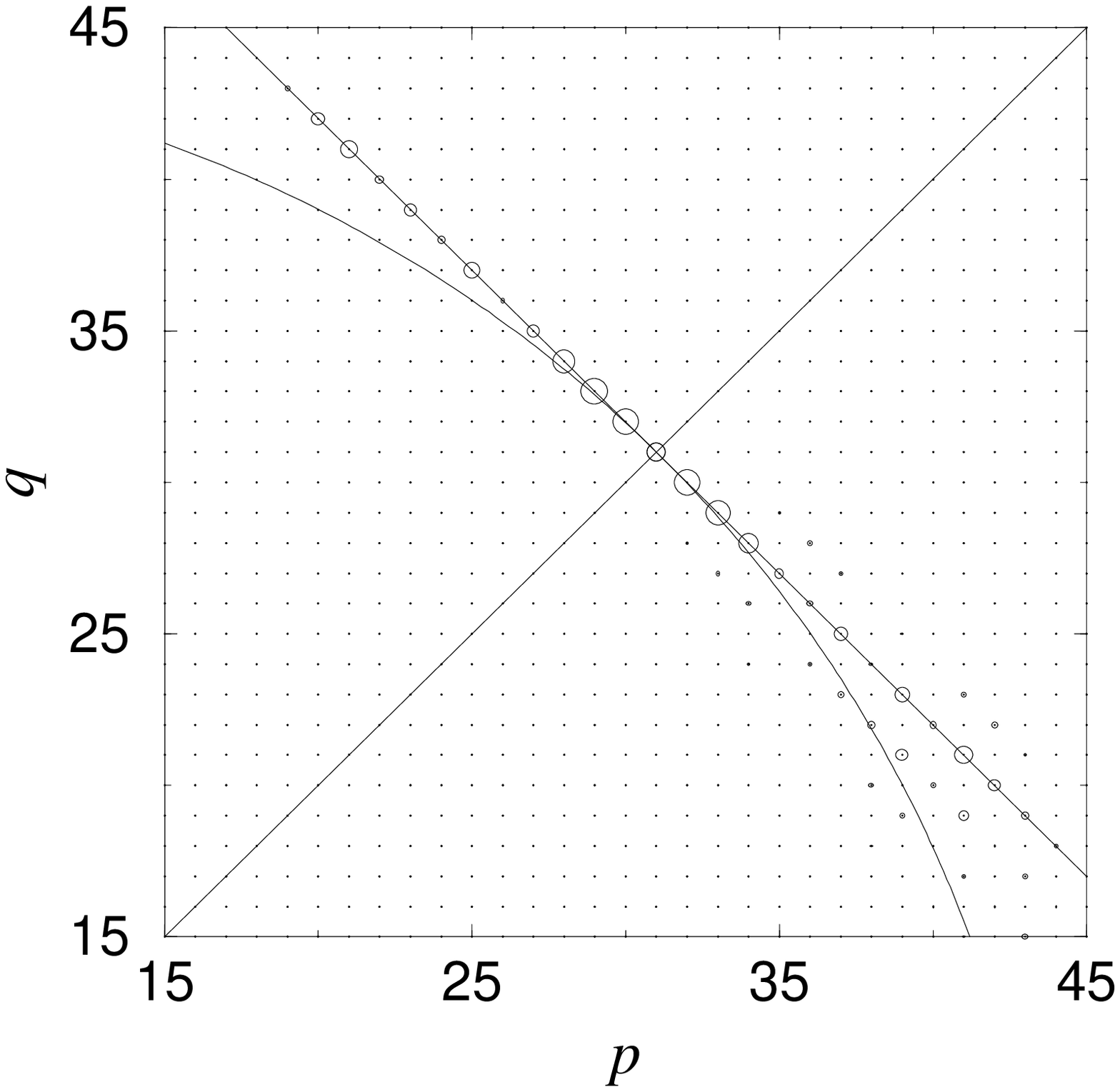,height=6cm,width=6cm,angle=270}}
{\vspace*{.13in}}
\caption[ty]{ The decomposition of the state of Fig. 6 in the $B=0$ basis states.
These are $\sin \pi p(x+\frac 12)\sin \pi q(y+\frac 12)$ symmetrized and
normalized.}
\end{figure}

While much of the above discussion can be carried over to the ABFL, in the
ideal case of a zero radius line there are strong diffraction effects which
limit the applicability of our theory. We therefore begin by defining the
alternative problem of a finite size flux line or tube. This is more
realistic if actual experiments are contemplated.

Let $\rho $ give the linear scale of the flux tube. We may think of the flux
as uniform inside a tube of this radius, or as being distributed in some
way, say as a gaussian, with $\rho $ giving the scale of the distribution.
[We actually used a square tube of side $\rho .$ Even more accurately, we
used four quarter strength tubes symmetrically located which allows a
symmetry reduction in the numerics. In the approximation of our theory, this
gives essentially the same result as a single circular tube. The field
inside a single flux tube is $B_0=\phi /4\rho ^2$ and $\phi $ is the total
flux.] The typical angular deflection suffered by a classical particle
traversing this field is 
\begin{equation}
\delta \theta \sim (\phi /\phi _0)/k\rho .  \label{deltheta}
\end{equation}
To avoid diffraction we require $\delta \theta $ to be small. This can be
achieved, of course, if $k\rho $ is large, but that is not necessary. In the
numerical work shown, we take $\phi /\phi _0=0.1,$ and $\rho =0.01,$ while $%
k\geq 140.$ An alternative and equivalent condition is to insist that, on
the appropriately defined average\cite{BO}, the terms in the Hamiltonian
satisfy $\left\langle (eA/c)^2/2m\right\rangle $ $<<\left\langle e{\bf %
p\cdot A}/mc\right\rangle $.

The results depend on where the flux line is located. We consider first the
case that it is located at $x=0,$ $y=-\frac 12+a$, where $0\leq a\leq \frac 1%
2.$ Again, we consider states related to the (1,1) resonance. This leads as
before to Eq. (\ref{um}) but now we find a potential 
\begin{eqnarray}
V_{AB}(x) &=&-Bk/{\cal L};\;x\in [-a,a],  \nonumber \\
V_{AB}(x) &=&+Bk/{\cal L};\;x\in [-1-a,-1+a],  \nonumber  \label{Vabfl} \\
V_{AB}(x) &=&0;\,x\notin [-a,a]\cup [-1-a,-1+a]  \label{Vabfl}
\end{eqnarray}
for points in $[-\frac 32,\frac 12],$ and $V_{AB}$ is extended periodically
by $V_{AB}(x+2)=V_{AB}(x).$ For the flux line, $B=2\pi \phi /\phi _0.$

Actually, Eq. (\ref{Vabfl}) is for the ideal flux line. The ideal case has
step function jumps in the potential which are smoothed out at finite $\rho $%
. In other words, the sharp jump at $x=a$, is replaced by a smooth rise
beginning at $x=a-\rho $ and ending at $x=a+\rho .$ The exact shape of the
rise depends on the distribution of flux in the line. As long as the
transverse wavelengths in the solution of Eq. (\ref{um}) are long compared
with $\rho ,$ the finiteness of $\rho $ does not play a significant role in
our theory at this level of approximation.

For the ideal, zero radius ABFL, there are significant deviations from this
scenario. Indeed, most matrix elements of the ABFL perturbed Hamiltonian in
the unperturbed basis are infinite. However, it is a weak, logarithmic
infinity, and our theory seems to capture the main shape of the
wavefunction, although at the relatively low energies for which numerical
results are available there are significant corrections. We consider these
to be diffractive corrections, arising from a characteristic length shorter
than the wavelength.

It is clear that $Bk/{\cal L}$ is the important parameter in the UF case,
while for the ABFL, both $Bk/{\cal L}$ and $a$ are important. We begin with
the case $a=\frac 14,$ which has a `square well potential' of width $\frac 12
$ near $x=0.$ For sufficiently large $Bk/{\cal L},$ there will be `tight
binding' solutions approximately $u_m(x)=\cos (m+1)\pi x/2a,\,\left|
x\right| <a,$ and zero elsewhere. This expression holds for sufficiently
small even $m,$ and for odd $m$ the cosine is replaced by the sine. The
energy $E_m\approx -Bk/{\cal L}+\pi ^2(m+1)^2/4a^2.$

Fig. 12 is for the ABFL case, with $n=86,$ $m=0,$ $r=0$ and $a=\frac 14.$
Fig. 12(a) shows $u_0(x)\ $ and $u_0(-x-1)$, its extension into $x<-\frac 12%
, $ reflected. For these parameters, $u_0$ is not extremely localized, and
extends significantly outside $[-\frac 12,\frac 12].$ The remaining plots
give $\left| \psi _m\right| =\left| \partial \Psi /\partial n\right| .$ Fig.
12(b) plots Eq. (\ref{Psi}) and 12(d) is from diagonalization in the limited
basis of the `degenerate' states. Clearly, these two approximations are
nearly the same. Fig. 12(c) and 12(e) are obtained by numerical
diagonalization in the complete unperturbed basis, for the finite size flux
tube and the ideal flux line respectively. Clearly, for these parameters,
there is significant diffraction from the flux tube. It is somewhat less
than for the zero radius line but the two patterns have some resemblance.
Diffraction evidently modifies the interference between different parts of
the wavefunction in an irregular way. The overall shape of $\left| \partial
\Psi /\partial n\right| $ is well predicted by the theoretical $u_0$. To
give a sense of the irregularity of diffraction effects on the wave
function, we show in Fig. 12(f) the function for $n=70,$ a somewhat lower
energy. In this case, according to theory, the $u_0$ is essentially the
same, and the interference fringes have a somewhat longer wavelength, on the
average. The diffractive effects are quite different in detail, however.

It is clear from this result that the ABFL localizes the particle, and to do
so, it must of course exert a force. That a flux line can exert a transverse
force is now well established\cite{shelankov}.

In Fig. 13 we show currents from a state with $a=\frac 12,$ i.e. the much
studied case with the flux line at the center of the square\cite{sieberdif}.
Again, $m=0,$ $r=0,$ and $n=$ $82$. The upper part of the figure shows
numerical streamlines for a corner of the square. Although the state is
spatially not well localized, it does have a strongly `paramagnetic' current
structure. By paramagnetic we mean $dE_m/d\phi <0$.

The lower part shows the theoretical current density $j_y(x)$ for $%
y=0,\,x\in [-\frac 12,0],$ that is, the current density along the upper edge
of the upper figure. A simple approximate formula for $j_y$ is $%
j_y(x)\propto -\left( \cos n\pi x/2\right) ^2\left[ u_m\left( -x-\frac 12%
\right) ^2-u_m\left( x-\frac 12\right) ^2\right] $ which gives double zeroes
of the current at equally spaced $x=(2l+1)/n.$ In this case $j_y$ is
negative and the factor depending on $u_m$ has no zeroes except at $x=0,$
for $m=0.$

The maxima are at $j_y=0$; $j_y\leq 0.$ Theory and numerics closely agree on
the period and shape of the oscillations, but the lower envelope of the two
differ. We mark with horizontal bars the lower envelope of the numerical
calculation, which is considerably more irregular than given by our theory.
This structure of zeroes of $j_x$ is a consequence of a symmetry, namely $%
y\leftrightarrow -y$, together with complex conjugation. This means that $%
\Psi (x,0)$ is real and therefore generically will have zeroes as a function
of $x.$

\section{Summary}

\subsection{Extensions of the results}

We have shown how to classify and find eigenstates of a charged particle in
a square billiard subjected to a magnetic flux which is classically weak. It
is not necessarily weak quantally, however, and large remarkable changes in
the wave functions are found which sometimes significantly localize the wave
function. We used two basic flux configurations: a uniform field and an
Aharonov-Bohm flux line. The placement of the ABFL is important. Of lesser
importance is any finite radius to the line, at least in the range of
parameters we use. We exhibited some of the wavefunctions in several forms
and sequences, concentrating on wavefunctions connected with the $(1,1)$
resonance, that is, connected with periodic orbits of the flux free square
whose velocities make angles of 45$^{\circ }$ with the coordinate axes.

These results can be readily generalized to higher order resonances. For a
given energy the effects diminish quite rapidly as the order increases, but
in principle, at sufficiently high energy any given resonance can show
strong magnetic effects. We give a few results in Appendix B.

\begin{figure}[tbp]
{\hspace*{0.2cm}\psfig{figure=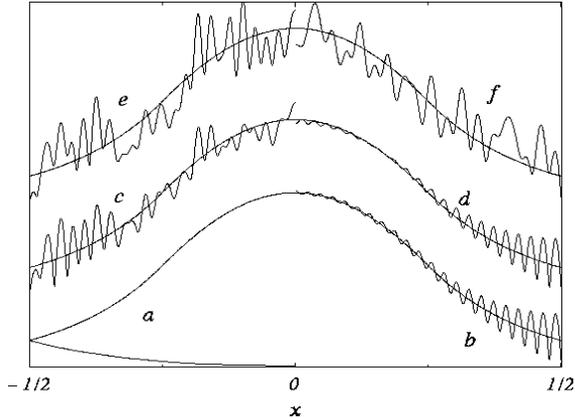,height=7.5cm,width=5.5cm,angle=270}}
{\vspace*{.13in}}
\caption[ty]{ ABFL case, $a=\frac 14,$ $\phi =$ $0.1\phi _0$, $\rho =0.01$, $n=$ $%
86$, $m=0.$ a) Upper curve, $u_0(x),$ lower curve, $u(-1-x).$ This is
repeated in the other pictures. The remaining plots are of $\left| \partial
\Psi /\partial n\right| .$ b) is the theoretical result from Eq. (\ref{Psi}%
), d) is from diagonalization in the `degenerate' basis of Section V (E),
and c) is from full numerical diagonalization. The interference oscillations
scale as $u(-1-x).$ e) and f) are numerical results for the ideal single
ABFL ($\rho =0$), with $n=82,$ and $n=70.$ The additional detailed structure
is attributed to diffraction, but the overall scale is given well by our
theory.}
\end{figure}

One can extend these results to integrable systems other than the square
billiard, and to other flux configurations. There are other billiard shapes,
such as the rectangle, certain triangles, the circle and the ellipse. One
may also study `soft billiards', e.g. a confining potential of the form $%
U(x,y)=U_1(x)+U_2(y)$ where $U_i$ is some sort of anharmonic potential. Soft
billiards are more accurate representations of mesoscopic systems than are
hard wall billiards, but the additional theoretical effort they require is
not usually made. A practical difficulty, although not one of principle, is
that it may be necessary to resort to action-angle variables, and it may be
tedious to find the periodic orbits of the flux-free system, if some tiling
trick cannot be used.

In any case, the first steps of the program are to choose a convenient
surface of section, choose the resonance of interest, find the AB phase $%
\Phi (x,x^{\prime }),$ and the effective potential $V(x).$ Much insight can
be gained at this level.

A much larger class of systems solvable by this technique are {\em nearly}
integrable systems subjected to a flux. For example, one could start from a
nearly square trapezoid. Then one would have a double perturbation of the
square, one from the flux, the other from the change of shape. Each of these
perturbations can have big quantum effects, and the combination of the two
can be quite different from each one separately, especially since one breaks
time reversal invariance and the other doesn't. Again, finding the effective
potential is key to understanding the qualitative results.

\subsection{Possible experiments}

There are possible experiments and even interesting devices which might be
made, if time reversal can be broken. Having unusual wavefunctions suggests
some of the possibilities. Probes of the system will depend on whether the
wavefunction is large or small at the position of the probe.

\begin{figure}[tbp]
{\hspace*{0.2cm}\psfig{figure=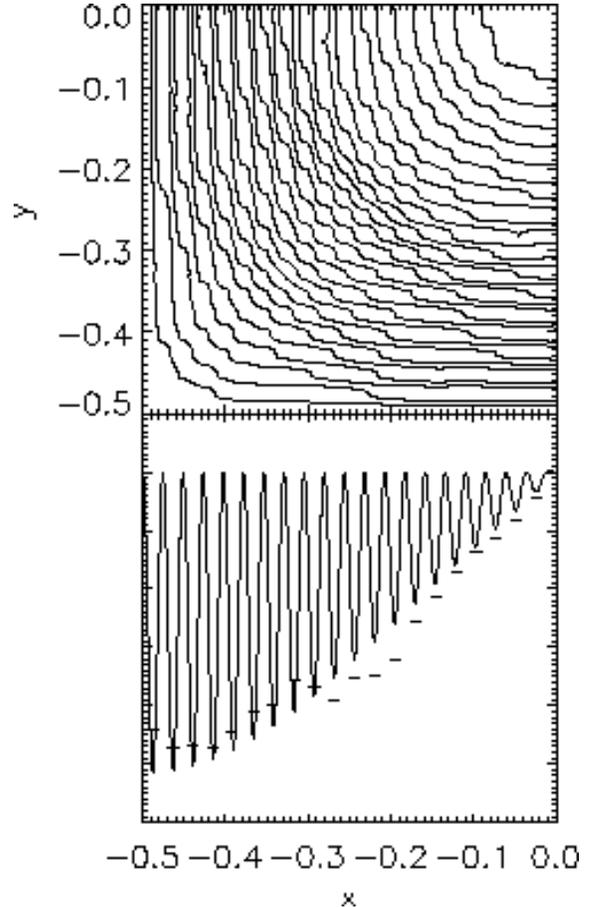,height=12cm,width=7.5cm,angle=0}}
{\vspace*{.13in}}
\caption[ty]{ Ideal zero radius flux line at the square center. Upper figure,  
numerical current streamlines for a quarter of the square. Lower figure,
theoretical current density $j_y(x,0)$. Dashes indicate the numerical minima.}
\end{figure}

Mesoscopic systems are of great current interest. In this case, electrons
are the particles and there is a magnetic flux. The system of L\'evy {\em et
al}.\cite{Levy} consisting of isolated metallic `squares' is a case of this
type, but one can imagine leads weakly connected to a metallic square whose
coupling depends on the shape of the wavefunction at the Fermi level and the
position of the lead.

Another type of mesoscopic system is formed of surface electrons in
`corrals' on a metallic surface\cite{corrals}. The wave functions of such
electrons can be probed with atomic accuracy with a scanning tunnelling
microscope. Achieving an appropriate parameter range will be difficult, but
perhaps not impossible. The corrals are leaky and do not really confine the
electrons to their interior, but in many cases that idea seems to work, at
least qualitatively.

Another kind of system is the shallow square or rectangular container of
liquid, which is vibrated to set up standing waves. Instead of a flux, the
system can be rotated with a uniform angular velocity. This differs from the
uniform magnetic field in that the $A^2$ is absent from the Hamiltonian, but
since under our assumptions that term can be neglected anyway, the same kind
of results are expected. One can also introduce a nonideal ABFL. This is
done by making a small hole in the tank and allowing the water to flow out
with a certain vorticity. Experiments on such a system have been performed%
\cite{berryv}, although scattering rather than eigenstates was studied. Such
an experiment would certainly have pedagogical value, and if the experiment
is done, it may in fact be the first detailed observation of a nontrivial
persistent current state.

Still another system is the thin square microwave cavity. Here the `quantum'
waves are microwaves. There is no quantity directly equivalent to the
magnetic field. However, one can replace part of the boundary by a ferrite
strip\cite{anlage}, say between $\left[ -a,a\right] $ on one side of the
cavity. The phase shift of the microwave upon reflection from the ferrite
depends on the direction of magnetization of the ferrite and the direction
of incidence of the microwave. This gives as an effective potential exactly
that of Eq. (\ref{Vabfl}). There is then a localized eigenstate circulating
the cavity in one direction, but not the other. A similar situation from the
point of view of diffraction and effective potential, but maintaining time
reversal invariance, is the `step' billiard\cite{step}. This is a square
billiard of side $L$ say, with one side moved down by $\delta L=\epsilon L$
for $\left| x\right| <a.$ The perturbed $(1,0)$ resonance states `see' a
square well effective potential to first approximation. If $k\delta L$ is
small, but $k\sqrt{L\delta L}\geq 1,$ our theory works, gives nontrivial
localization, and is similar to the flux line with small $\phi /\phi _0.$
The range $k\delta L=\pi ,$ i.e. $\delta L$ a half wavelength, gives no
phase shift for normal incidence and is similar to an ABFL with $\phi /\phi
_0=1.$ Experiments on this system could be carried out in several contexts.

\subsection{Conclusions}

We have solved a characteristic example of quantum states in a weakly
perturbed integrable system. The states and their energies can be classified
and found to good approximation. The wavefunctions are quite nontrivial and
interesting, as compared with the states of the integrable system. The
states of a hard chaotic system are also not individually interesting,
except for some weak `scars', requiring resort to statistical studies and
averages over many states in such systems.

Our technique applies to a very large class of systems, which includes
experimental systems. The particular case we emphasize breaks time reversal
invariance, and the eigenstates have persistent currents. As far as we know,
these are the first published nontrivial examples such states.

\section{Acknowledgments}

Supported in part by the United States NSF grant DMR-9625549 and United
States-Israel Binational Science Foundation, grant 99800319. R.N. was
partially supported by the NSF grant DMR98-70681 and the University of
Kentucky. We thank Prof. Director Peter Fulde for hospitality at the
Max-Planck-Institut f\"ur Physik komplexer Systeme in Dresden, where some of
this work was done. R.N. and O.Z. thank Dr. R. Seiler for hospitality at the
SFB 288 ``Differentialgeometrie und Quantenphysik'', TU Berlin. O.Z. thanks
Dr. F. Haake for hospitality at the Universit\"at GH Essen. \appendix 

\section{Two dimensional wavefunctions}

In this Appendix we find the two dimensional wavefunction. If a solution to $%
\psi =T\psi $ is known, the wavefunction in the square can be found by
evaluating the integral over the surface of section\cite{bogolss}. 
\begin{equation}
\Psi ({\bf r})=\int G({\bf r},x^{\prime };E)\psi (x^{\prime })dx^{\prime }.
\label{intGpsi}
\end{equation}
In this case 
\begin{eqnarray}
G({\bf r},x^{\prime };E) &=&\sum_{\text{cl. tr.}}\frac 12\left| \frac 1{2\pi
k}\frac{\partial ^2S}{\partial x^{\prime }\partial \ell _{\perp }}\right|
^{1/2}  \nonumber \\
&&\times \exp \left[ iS({\bf r},x^{\prime };E)-i\frac \pi 2\nu \right]
\label{Green}
\end{eqnarray}
in our units. Here the sum is over the (unperturbed) classical trajectories
going from a point $x^{\prime }$ on the surface of section to a point ${\bf r%
}$ inside the square, $S({\bf r},x^{\prime };E)$ is the reduced action for
this trajectory, $\ell _{\perp }$ is the direction perpendicular to the
trajectory at point ${\bf r}$, and $\nu $ is the Maslov index. In a single
square scheme we take the surface of section to be the lower side of the
square $-1/2\leq x^{\prime }\leq 1/2$, $y^{\prime }=-1/2$. Then $\psi
(x^{\prime })$ is a linear combination of functions $\psi _I$ and $\psi
_{II} $ described in the text such that $\psi (\pm 1/2)=0$ (there should be
no flow through the ends). It follows then that $u_m(x^{\prime
}+2)=(-1)^ru_m(x^{\prime })$ and 
\begin{equation}
\psi (x^{\prime })=e^{i\kappa x^{\prime }}u_m(x^{\prime })+i^re^{-i\kappa
x^{\prime }}u_m(x^{\prime }-1)  \label{psiIpsiII}
\end{equation}
with $r$ {\it defined }by Eq. (\ref{groupr}). Here we have also used a zero
field quantization condition $\kappa =\pi n/2$ and the property $%
u_m(x^{\prime })=(-1)^mu_m(-x^{\prime })$. Note that in the surface of
section picture $r$ does not have a direct interpretation as a
representation label.

We evaluate the integral (\ref{intGpsi}) in the stationary phase
approximation. As we shall see, only the 45$^{\circ }$ orbits survive. For a
given point ${\bf r}=(x,y)$ there are four such orbits: two starting at $%
x_1^{\prime }$ and two at $x_2^{\prime }$ (Fig. 14). Each orbit gives one
term in Eq. (\ref{Psi}). Note that the orbits with the positive (negative) $%
x $-projection of the momentum at $x^{\prime }$ are generated by the first
(second) term in Eq. (\ref{psiIpsiII}).

\begin{figure}[tbp]
{\hspace*{0.2cm}\psfig{figure=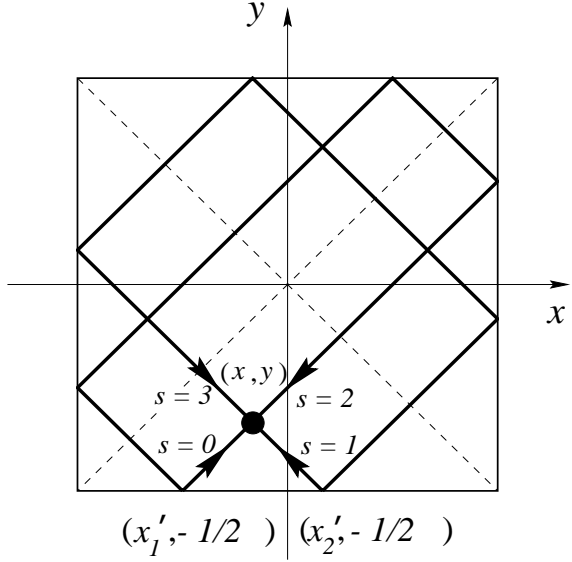,height=7.5cm,width=7.5cm,angle=0}}
{\vspace*{.13in}}
\caption[ty]{ A diagram showing how the full wavefunction $\Psi (x,y)$ is 
constructed from the surface of section wavefunction. Two points on the
surface of section, $x_1^{\prime },$ $x_2^{\prime }$ dominate in the sense
of stationary phase, and there are two paths from each of these points to
the point $(x,y)$.}
\end{figure}

Although the following calculation depends in which sector inside the square
the point lies, the final result is independent of sector. So for
definitiveness we assume $y<x<-y$, that is the point lies below the
diagonals $y=x$ and $y=-x$. We have numbered the orbits by $s=0,\ldots ,3$
according to the terms in Eq. (\ref{Psi}) they represent. For $s=0$ orbit,
expressing the action in terms of the distance, $S({\bf r},x^{\prime };E)=kL(%
{\bf r},x^{\prime })\equiv k\sqrt{(x-x^{\prime })^2+(y+1/2)^2}$, we find the
prefactor in Eq. (\ref{Green}) 
\begin{equation}
\left| \frac 1{8\pi k}\frac{\partial ^2S}{\partial x^{\prime }\partial \ell
_{\perp }}\right| ^{1/2}=\left( \frac 1{8\sqrt{2}\pi L}\right) ^{1/2}.
\end{equation}
The stationary point in the integral (\ref{intGpsi}) is determined by the
exponents in Eq. (\ref{Green}) and in the first term of Eq. (\ref{psiIpsiII}%
). It is $x_1^{\prime }=x-y-1/2$, as in a 45$^{\circ }$ orbit. The $s=0$
wavefunction is then 
\begin{equation}
\Psi _0(x,y)=e^{i\kappa (x+y)}u_m\left( x-y-\frac 12\right)  \label{Psi0}
\end{equation}
dropping a constant factor $\kappa ^{-1/2}e^{i(\kappa /2-\pi /4)}/2$. The
terms with $s=1,2,3$ can be obtained in the same manner with an appropriate
addition of the Maslov phase for each bounce.

If ${\bf r}$ is located in a different sector of the square, the results
remain the same if proper account is made of the Maslov phases. The labeling
of the orbits is invariant if done by the rule: $s=0$ trajectory arrives to $%
{\bf r}$ from the South-West, $s=1$ from SE, $s=2$ from NE, and $s=3$ from
NW.

It is worth noticing that Eqs. (\ref{Psi0}) or (\ref{Psi}) are valid only up
to the square root order of the small parameter $\epsilon =B/k$ of the
perturbation theory behind our work. If we wish to obtain the results valid
to the first order of $\epsilon $ we should (a) use a better approximation
for $u_m(x^{\prime })$\cite{pnz}, (b) add the vector potential term to the
action $S({\bf r},x^{\prime };E)$, and (c) find a correction to the
stationary point due to $u_m(x^{\prime })$.

Alternatively, the above calculation can be carried out in the repeated
square scheme defined in Sec. III. The surface of section is then the line $%
y=-1/2$ (identified with $y=3/2$). We can restrict ourselves to the
trajectories with the positive $x$- and $y$-projections of the momentum in
this extended manifold. Then only $\psi _I(x^{\prime })$ is needed to
generate a complete two-dimensional wavefunction. In order for $\psi _I$ to
have a period 2, the condition $u_m(x^{\prime }+2)=(-1)^ru_m(x^{\prime })$
must be satisfied. A point ${\bf r}$ in the original square will be
represented by its images in four (up to a translation by $\Delta x=2$)
domains in the manifold. Each of these images is connected by {\it one} 45$%
^{\circ }$ orbit to the surface of section. These four orbits generate the
four terms $s=0,\ldots ,3$ in Eq. (\ref{Psi}) as follows: $s=0$ is produced
by the point ${\bf r}$ in the original domain, $s=1,3,2$ by its reflection
about the right side of the square, upper side of the square, and the
composition of both, respectively. There are no Maslov phases in Eq. (\ref
{Green}) in this case. However the terms $s=1,3$, which are obtained by the
odd number of reflections, should be summed with an additional minus sign.
Indeed, the Jacobian of the coordinate transformation from the original
square to the extended manifold is singular on the boundary. So, when the
wavefunction on the manifold is folded back to the physical domain a phase
difference $\pi $ will be accumulated between each pair of domains related
by one reflection. Of course, this phase is analogous to the Maslov phase
for an impenetrable wall.

\section{Higher resonances}

For completeness, we give some results for higher resonances, which can be
labelled with relatively prime integers, $p,q.$ These resonances are closely
related to states of a rectangle of side $1/p,$ $1/q.$ Obviously, for the
square, the energies for $p,q$ are the same as those for $q,p.$ The
corresponding states are also the same after a $90^{\circ }$ rotation.
However, for $\left| p-q\right| /\left| p+q\right| $ not too small, the $p,q$
states and the $q,p$ states are different and are nearly uncoupled and there
is a nearly doubly degenerate set of energy levels. The splitting of the
exact levels, which are combinations of $pq$ and $qp$ that are eigenstates
of the rotation operator ${\cal R}$, can be estimated by using an analog of
`chaos assisted tunnelling'.

The classical periodic orbits make $q\,$bounces from the $x$ sides and $p$
bounces from the $y$ sides. These resonant orbits all have the same length $%
{\cal L}_{pq}=2\sqrt{p^2+q^2}.$ The maximal directed area enclosed by an
orbit is $\pm 1/2pq.$ The energy is $k_{nm}^2$ where 
\begin{equation}
k_{nm}^2\approx \left( \frac{2\pi n}{{\cal L}_{pq}}\right) ^2+\left[ 1+\frac{%
p^2}{q^2}\right] E_m^{(q)}  \label{kpq}
\end{equation}
and $n$ is an integer. The first term is much larger than the second. Again,
we have the approximate quantization of $n$ wavelengths in an orbit length.

The potential $V_{pq}(x)$ depends on $q$ being associated with the $x$%
-direction, thus the notation $E_m^{(q)}$. This `transverse' contribution to
the energy is of course symmetric in $p$ and $q,$ that is, $%
q^2E_m^{(p)}=p^2E_m^{(q)}.$ The potential is 
\begin{equation}
V_{pq}(x)=\frac qp\left[ \frac{{\cal L}_{11}}{{\cal L}_{pq}}\right]
^3V\left[ q\left( x+\frac 12\right) -\frac 12\right]  \label{Vpq}
\end{equation}
where $V$ is given by Eq. (\ref{Vuf}). $V_{pq}$ has period $2/q$, rather
than $2.$ The boundary condition on the eigenstates is $u(x-2)=e^{i\beta
}u(x)$ where $\beta =2\pi 
%TCIMACRO{\limfunc{frac}}
%BeginExpansion
\mathop{\rm frac}
%EndExpansion
\left[ pn/\left( p^2+q^2\right) \right] $ and $%
%TCIMACRO{\limfunc{frac}}
%BeginExpansion
\mathop{\rm frac}
%EndExpansion
$ indicates the fractional part. The potential can be regarded as weak if $%
kB<<pq{\cal L}_{pq}^3.$ If this condition is satisfied the $pq$ resonance
can be ignored.

Similar but somewhat more complex results can be obtained for rectangles and
certain triangles.

\end{document}